# Minimum model and its theoretical analysis for superconducting materials with BiS$_2$ layers

K. Suzuki[*], H. Usui, K. Kuroki

*Department of Engineering Science, The University of Electro-Communications, 1-5-1 Chofugaoka, Chofu, 182-8585, Japan*

**Abstract**

We perform first principles band calculation of the newly discovered superconductor LaO$_{1-x}$F$_x$BiS$_2$, and study the lattice structure and the fluorine doping dependence of the gap between the valence and conduction bands. We find that the distance between La and S as well as the fluorine doping significantly affects the band gap. On the other hand, the four orbital model of the BiS$_2$ layer shows that the lattice structure does not affect this portion of the band. Still, the band gap can affect the carrier concentration in the case of light electron doping, which in turn should affect the transport properties.

## 1. Introduction

Recently, Mizuguchi *et al*. have discovered new superconducting materials possessing BiS$_2$ layers, where the Bi and S atoms are aligned alternatively on a square lattice [1,2]. So far there are two kinds of materials; one is Bi$_4$O$_4$S$_3$ which contains Bi in the blocking layer, and the other is $Re$O$_{1-x}$F$_x$BiS$_2$ ($Re$=Rare earth) [2-5] with a maximum $T_c$=10.6K in LaO$_{0.5}$F$_{0.5}$BiS$_2$.

Band structure calculations suggest that the conduction bands mainly consist of Bi 6$p_x$, 6$p_y$ orbitals within the BiS$_2$ layer hybridized with S 3$p$ orbitals [1,6]. Below the band gap lies the valence band, whose top mainly consists of O 2$p$ orbitals for the mother compound. The S 3$p$ bands lie at somewhat lower energies in the mother compound, and these bands have been extracted in ref. [8] by constructing a four orbital model of the BiS$_2$ layer of LaOBiS$_2$. Some theoretical studies have investigated the pairing mechanism, where electron-phonon [7-9] or electron correlation effects [10] play important roles.

There have also been some interesting experimental observations other than the superconductivity itself. Resistivity of Bi$_4$O$_4$S$_3$ exhibits a metallic behaviour, while LaO$_{0.5}$F$_{0.5}$BiS$_2$ shows a semiconducting behaviour at high pressures [11]. There, the authors have considered a possibility that the carrier concentration may not be so large as expected from the nominal fluorine content, and the carrier doping may occur due to self doping. For LaO$_{1-x}$F$_x$BiS$_2$, it has been shown in ref. [12] that resistivity decreases as $x$ is increased up to $x$~0.5, but then exhibits a semiconducting behaviour for larger fluorine content. For CeOBiS$_2$ on the other hand, it has been shown that the mother compound is a bad metal, but shows a semiconducting behaviour with fluorine doping [3].

When electrons are significantly doped as expected from the nominal fluorine content of $x$~0.5, then there is a possibility of some kind of instability giving rise to a gap [3]. On the other hand, in the lightly doped case (including the possibility that the electrons are not doped so much as expected from the nominal fluorine content), the size of the band gap between the conduction and the valence bands should play an important role in determining the actual carrier concentration.

In the present paper, we discuss the lattice structure and the fluorine doping dependence of the band gap size. We focus on LaOBiS$_2$, because it has a simple lattice structure compared to Bi$_4$O$_4$S$_3$. We perform first principles band calculation of LaOBiS$_2$ and LaO$_{0.5}$F$_{0.5}$BiS$_2$ using several lattice structures, and discuss the effect of lattice structure and the doping dependence on the band gap. We also obtain the four orbital model which consider Bi 6 $p_x$, $p_y$ and S 3 $p_x$, $p_y$ to see how the lattice structure affects these main bands.

---

*Corresponding author. Tel.: +81-42-443-5559; +81-42-443-5563.
*E-mail address*: suzu@vivace.e-one.uec.ac.jp.

## 2. Crystal structure dependence of band gap size

First, we perform first principle band calculation of the mother compound LaOBiS$_2$ (without F doping) using the Wien2k package [13] and adopting two experimentally determined lattice structures, i.e., that of LaOBiS$_2$ given in ref. [14], and that of LaO$_{0.5}$F$_{0.5}$BiS$_2$ given in ref. [1]. (The latter is a hypothetical structure for the mother compound). Here we take $RK_{max}$=7, 512-kpoints, and adopt GGA-PBE exchange correlation functional [15]. The comparison is given in Fig. 1 (a) (b), which shows that the band gap size largely depends on the lattice structure. In particular, the gap nearly closes for the lattice structure of LaO$_{0.5}$F$_{0.5}$BiS$_2$. Next, we perform band calculation for the fluorine doped LaO$_{0.5}$F$_{0.5}$BiS$_2$, using the virtual crystal approximation, and again adopting the two lattice structures. The results given in Fig. 1 (c) (d) shows that the top of the valence bands around the Γ point, which is present for LaOBiS$_2$ and mainly originates from O 2 $p$ orbitals, sink for the F doped case, and in turn, the bands around the X point, which consist of mainly of S 3 $p$ orbitals (hybridized with Bi 6$p$) surface as the valence band top. This modification of the band structure by fluorine doping significantly enlarges the band gap. Here again the lattice structure affects the band structure, but not the band gap itself. So the present results show that both the lattice structure and the fluorine doping affects the band gap, but in a different manner.

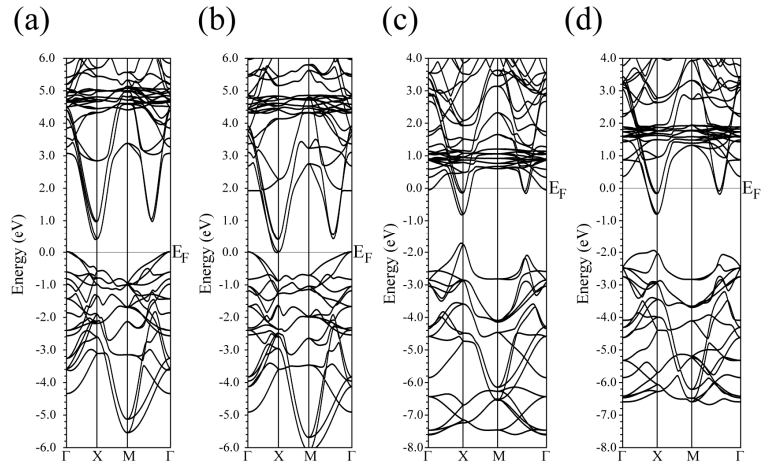

Fig. 1: Band structure of LaOBiS$_2$ obtained using the lattice structures of (a) LaOBiS$_2$ [14] and (b) LaO$_{0.5}$F$_{0.5}$BiS$_2$ [1], and that of LaO$_{0.5}$F$_{0.5}$BiS$_2$ adopting virtual crystal approximation and using the lattice structures of (c) LaOBiS$_2$ and (d) LaO$_{0.5}$F$_{0.5}$BiS$_2$.

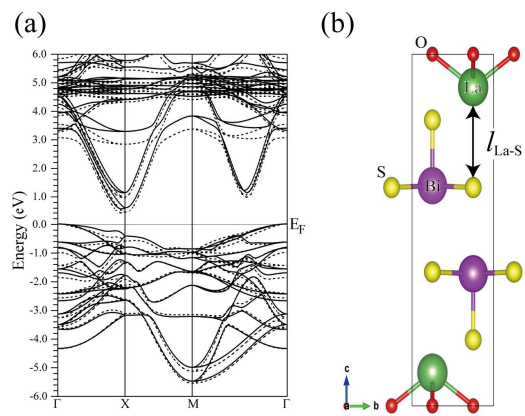

Fig. 2: (a) Band structure of LaOBiS$_2$ adopting the *c*-axis lattice constant of LaO$_{0.5}$F$_{0.5}$BiS$_2$ [1] while fixing other parameters at the original values [14] (solid lines). The original band structure is also shown for comparison (dashed lines) (b) The lattice structure of LaOBiS$_2$ and the definition of $l_{La-S}$.

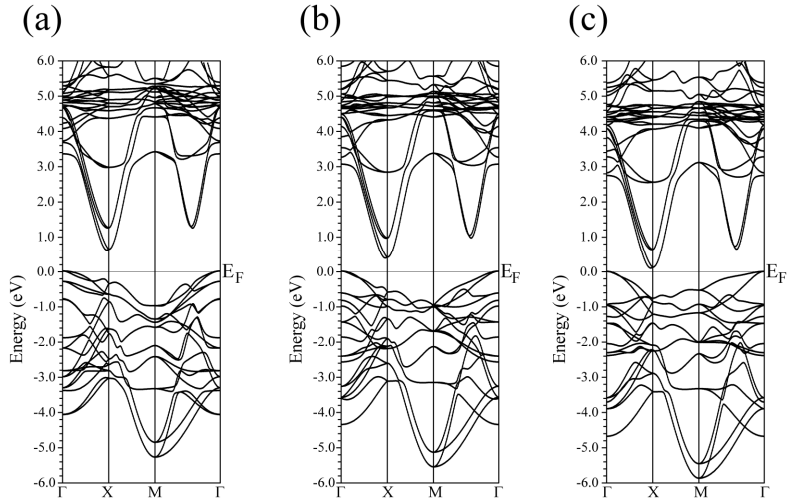

Fig. 3: $l_{La-S}$ dependence of the band structure of LaOBiS$_2$. $l_{La-S}$= (a) 4.11 Å, (b) 3.92 Å (original) and (c) 3.83 Å.

### 3. Hypothetical lattice structures

To understand the essence of the lattice structure sensitivity of the band gap, we focus on the non-doped LaOBiS$_2$ and consider hypothetical lattice structures. One of the large differences between the lattice structures of ref. [1] and [14] is the *c*-axis lattice constant. In Fig. 2 (a), we show the band structure of LaOBiS$_2$, where we adopt only the *c*-axis constant given in ref. [1], while all the other parameters are as in ref. [14]. This shows that the *c*-axis length itself barely affects the band gap.

In order to pin down the origin of the lattice structure sensitivity, we have actually considered various hypothetical lattices with varied lattice parameters, and have found that the LaO layers play an important role. Here we introduce the La-S length $l_{La-S}$ as shown in Fig. 2 (b), and vary this length hypothetically by moving the La position along the *c*-axis. Figs. 3 (a)-(c) show comparison of the band structure of LaOBiS$_2$ for various $l_{La-S}$. They show that larger $l_{La-S}$ expands the band gap. These results suggest that the position of the atoms outside of the BiS$_2$ layer affects the band gap size, so that it can affect the transport properties when the doped carrier concentration is small.

In the present study, we have not included the spin-orbit (SO) coupling in the calculation, but a calculation which includes the SO coupling shows that the band gap tends to be smaller. Therefore, there is a strong possibility that the band gap closes for the non-doped or lightly doped materials depending on the lattice structure, and this can result in self-doping of carriers. This may be relevant to some of the experimental observations mentioned in the Introduction.

### 4. Minimal model

We have seen that the band gap is affected by the lattice structure as well as fluorine doping. In this section, we address the question of whether the lattice structure significantly affects the portion of the bands that is *directly* responsible for the superconductivity. The minimal model that extracts the relevant band structure of the BiS$_2$ layer has been obtained in ref. [6]. Using the same formalism that exploits the maximally localized Wannier orbitals [16], we construct a minimal model for the hypothetical lattice structure in which $l_{La-S}$ is reduced to 3.83 Å. The result is shown in Fig. 4 in comparison with that for the original lattice structure. It can be seen that the band structure is barely affected by the lattice structure. Still, even if the main band structure of the BiS$_2$ layer itself is unaffected, the band gap can affect the carrier concentration in the case of light electron doping, which in turn should affect the transport and superconducting properties as mentioned above.

### 5. Conclusion

In conclusion, we have performed first principle band calculation of LaO$_{1-x}$F$_x$BiS$_2$ and discussed the lattice structure and the fluorine doping dependence of the band gap size. We find that the distance between La and S significantly affects the band gap. The four orbital model for the BiS$_2$ layer shows that the lattice structure does not affect this portion of the band. Still, the band gap can affect the carrier concentration in the case of light electron doping, which in turn should affect the transport and superconducting properties.

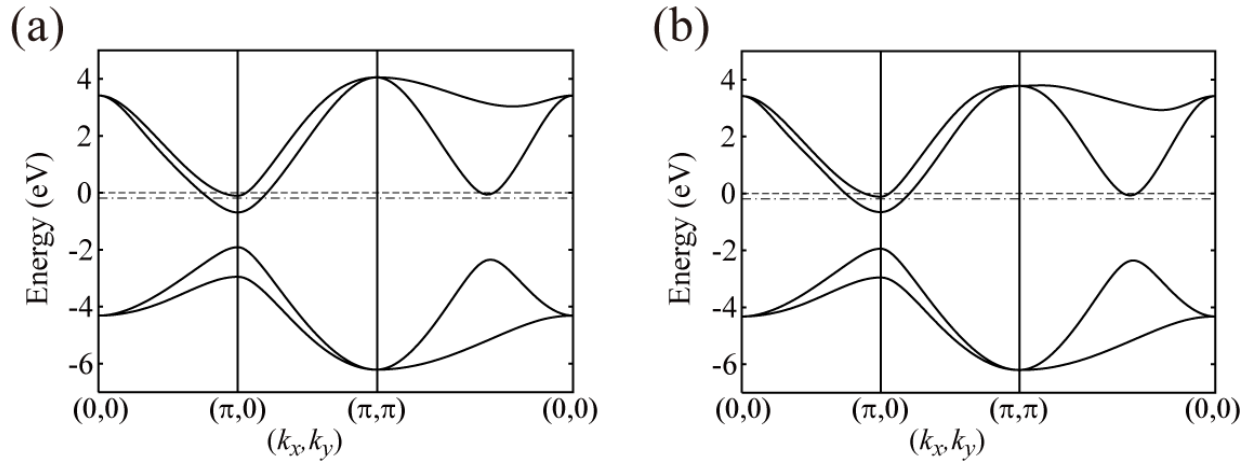

Fig. 4: Band structure of the four orbital model obtained for (a) the original LaOBiS$_2$ lattice structure [14] and (b) the hypothetical one with $l_{\text{La-S}}$=3.83. Dashed (dash-dotted) lines indicate the Fermi energy for the doping ratio of $x$=0.5 ($x$=0.25). E=0 corresponds to the Fermi energy of $x$=0.5.

## Acknowledgements

We would like to thank Y. Mizuguchi and collaborators in ref. [2] for valuable discussions.

## References

[1] Y. Mizuguchi, H. Fujihisa, Y. Gotoh, K. Suzuki, H. Usui, K. Kuroki *et al.*, Phys. Rev. B 86 (2012) 220510.
[2] Y. Mizuguchi, S. Demura, K. Deguchi, Y. Takano, H. Fujihisa, Y. Gotoh *et al.*, J. Phys. Soc. Jpn. 81 (2012) 114725.
[3] J. Xing, S. Li, X. Ding, H. Yang, H.-H. Wen, Phys. Rev. B 86 (2012) 214518.
[4] R. Jha, S. K. Singh, V. P. S. Awana, arXiv: 1208.5873.
[5] S. Demura, Y. Mizuguchi, K. Deguchi, H. Okazaki, H. Hara, T. Watanabe *et al.*, arXiv: 1207.5248.
[6] H. Usui, K. Suzuki, K. Kuroki, Phys. Rev. B 86 (2012) 220501.
[7] X. Wan, H.-C. Ding, S. Y. Savrasov, C.-G. Duan, arXiv: 1208.1807.
[8] T. Yildirim, arXiv: 1210.2418.
[9] B. Li, Z. W. Xing, G. Q. Huang, arXiv: 1210.1743.
[10] Y. Liang, X. Wu, W.-F. Tsai, J. Hu, arXiv: 1211.5435.
[11] H. Kotegawa, Y. Tomita, H. Tou, H. Izawa, Y. Mizuguchi, O. Miura *et al.*, J. Phys. Soc. Jpn. 81 (2012) 103702.
[12] K. Deguchi, Y. Mizuguchi, S. Demura, H. Hara, T. Watanabe, S. J. Denholme *et al.*, arXiv: 1209.3846.
[13] P. Blaha, K. Schwarz, G. K. H. Madsen, D. Kvasnicka, J. Luitz, *Wien2k: An Augmented Plane Wave + Local Orbitals Program for Calculating Crystal Properties* (Vienna University of Technology, Wien, 2001).
[14] V. S. Tanryverdiev, O. M. Aliev, Inorg. Mater. 31 (1995) 1361.
[15] J. P. Perdew, K. Burke, M. Ernzerhof, Phys. Rev. Lett. 77 (1996) 3865.
[16] N. Marzari, D. Vanderbilt, Phys. Rev. B 56 (1997) 12847; I. Souza, N. Marzari, D. Vanderbilt, *ibid.* 65 (2001) 035109. The Wannier functions are generated by the code developed by A. A. Mostofi, J. R. Yates, N. Marzari, I. Souza, D. Vanderbilt, (http://wannier.org/).